\documentclass[prd,twocolumn,preprintnumbers,nofootinbib,superscriptaddress]{revtex4-1}
\pdfoutput=1
\usepackage{cancel}

\RequirePackage[colorlinks=true
,urlcolor=blue
,anchorcolor=blue
,citecolor=blue
,filecolor=blue
,linkcolor=blue
,menucolor=blue
,linktocpage=true
,pdfproducer=medialab
,pdfa=true
]{hyperref}

\usepackage{amsmath,amssymb,amsthm,amsfonts,mathtools}
\usepackage{graphicx,tabularx}
\usepackage{color}
\usepackage{multirow}
\usepackage{comment}
\usepackage{enumitem}
\usepackage{physics}
\usepackage{cleveref}
\usepackage{slashed}
\usepackage[normalem]{ulem}
\usepackage{hyperref}

\usepackage{scalerel}
\setlist[description]{leftmargin=0.3cm}
\setlist[itemize]{leftmargin=0.5cm}

\newcommand{\be}{\begin{equation} \begin{aligned}}
\newcommand{\ee}{\end{aligned} \end{equation}}
\newcommand{\beqa}{\begin{eqnarray}}
\newcommand{\eeqa}{\end{eqnarray}}

\def\figureautorefname~#1\null{Fig.\,#1\null}
\def\tableautorefname~#1\null{Tab.\,#1\null}

\def\equationautorefname~#1\null{Eq.\,(#1)\null}

\RequirePackage[normalem]{ulem}

\crefname{section}{Sec.}{Secs.}
\crefname{figure}{Fig.}{Figs.}
\crefname{equation}{Eq.}{Eqs.}
\crefname{appendix}{Appendix}{Appendices}

\definecolor{kjkblue}{rgb}{0.39, 0.589, 0.6914}
\definecolor{darkgreen}{RGB}{0, 171, 148}

\def\Fermilab{Theoretical Physics Department, Fermilab, P.O. Box 500, Batavia, IL 60510, USA}
\def\IFUSP{Instituto de F\'isica, Universidade de S\~ao Paulo, C.P. 66.318, 05315-970 S\~ao Paulo, Brazil}

\begin{document}

\preprint{FERMILAB-PUB-24-0560-T, IPPP/24/59, UCI-HEP-TR-2024-15}

\title{Measuring the weak mixing angle at SBND}

\author{Gustavo F. S. Alves}
\email{gustavo.figueiredo.alves@usp.br}
\affiliation{\IFUSP}
\affiliation{\Fermilab}

\author{Antonio P. Ferreira}
\email{antoniopf99@usp.br}
\affiliation{\IFUSP}
\affiliation{\Fermilab}

\author{Shirley Weishi Li}
\email{shirley.li@uci.edu}
\affiliation{\Fermilab}
\affiliation{Department of Physics and Astronomy, University of California, Irvine, CA 92697}

\author{Pedro A.~N. Machado}
\email{pmachado@fnal.gov}
\affiliation{\Fermilab}

\author{Yuber F. Perez-Gonzalez}
\email{yuber.f.perez-gonzalez@durham.ac.uk}
\affiliation{Institute for Particle Physics Phenomenology, Durham University, South Road DH1 3EL, Durham, United Kingdom}

\date{September 11, 2024}

\begin{abstract}
The weak mixing angle provides a sensitive test of the Standard Model.
We study SBND's sensitivity to the weak mixing angle using neutrino-electron scattering events. We perform a detailed simulation, paying particular attention to background rejection and estimating the detector response. 
We find that SBND can provide a reasonable constraint on the weak mixing angle, achieving 8\% precision for $10^{21}$ protons on target, assuming an overall flux normalization uncertainty of 10\%. 
This result is superior to those of current neutrino experiments and is relatively competitive with other low-energy measurements.\footnote{The work and conclusions presented in this publication are not to be considered as results from the SBND collaboration.}
\end{abstract}

\maketitle
\flushbottom

\section{Introduction}
\label{sec:Introduction}

The weak mixing angle, $\sin^2\theta_W$, is a crucial parameter in the Standard Model (SM). 
Precise measurements of $\sin^2\theta_W$ across different energy scales---guided by quantum corrections---are essential for testing the electroweak sector and probing light, weakly coupled new physics.
Deviations from the SM-predicted running can indicate the presence of new particles or interactions~\cite{ParticleDataGroup:2022pth, Davoudiasl:2014kua}.

Neutrinos are a valuable tool for probing the gauge structure of the SM, as they interact only via electroweak interactions. 
However, measuring $\sin^2\theta_W$ with neutrinos is challenging. 
Neutrino-nucleus scattering is the typical signature in neutrino experiments due to its large cross section, but this channel is limited by theoretical uncertainties~\cite{Wilkinson:2022dyx} and cross section mis-modeling~\cite{Coyle:2022bwa}.
Current and future neutrino experiments operate at an energy range around $0.1-10$~GeV, where challenges related to cross section modeling are prominent.
For example, data from the MINERvA~\cite{MINERvA:2023kuz} and NOvA~\cite{NOvA:2022see} experiments reveal significant discrepancies from theoretical model predictions, highlighting the limitations of current models. 
Additionally, tuning these models to fit near-detector data complicates their use for performing precision tests~\cite{Coyle:2022bwa}.

Despite the theoretical challenges, the NuTeV experiment has provided the most competitive neutrino measurement of the weak mixing angle. 
By using the ratio of charged- to neutral-current neutrino-iron scattering cross sections at high energies, $E_\nu \sim 100$~GeV~\cite{NuTeV:2001whx}, NuTeV reported a measurement of $\sin^2\theta_W=0.2277\pm0.0016$, showing a $3\sigma$  discrepancy with the SM prediction based on LEP results. 
However, several factors not accounted for in the NuTeV analysis could help explain this discrepancy~\cite{ParticleDataGroup:2022pth}. These include strange-quark parton distribution functions (PDFs), isospin symmetry breaking in PDFs and splitting functions, and nuclear shadowing. 
The NuTeV result underscores the need for an alternative measurement of $\sin^2\theta_W$ in neutrino experiments that does not suffer from nuclear cross section modeling uncertainties.

Neutrino-electron scattering offers such an alternative.
This process has a well-understood cross section and reliable event reconstruction in detectors. 
The primary drawback is the small cross section, resulting in low statistics, with neutrino-nucleus interactions serving as the main background. 
The Short-Baseline Near Detector (SBND) experiment~\cite{MicroBooNE:2015bmn} is well-suited to leverage this channel, as it will collect a substantial amount of data to mitigate statistical limitations. 
Furthermore, as a liquid argon time projection chamber, SBND is poised to reject backgrounds effectively using cuts on hadronic activities and angular distributions of electromagnetic showers~\cite{deGouvea:2019wav}.
Another advantage of SBND is its proximity to the Booster Neutrino Beam (BNB) target, which makes it sensitive to the beam's angular spread.
This spread, primarily due to the kinematics of charged mesons and their decay products in the beam, can be utilized to reduce systematic uncertainties through the SBND-PRISM technique~\cite{delTutto2021SBND}.

In this paper, we analyze SBND's sensitivity to the weak mixing angle. 
We assess the statistical limitations of measuring $\sin^2\theta_W$ at SBND and estimate the exposure required for a measurement competitive with atomic parity violation and other methods. 
We also investigate the role of SBND-PRISM in mitigating flux uncertainties. 
This paper is organized as follows. 
In Sec.~\ref{sec:nu_e_scattering}, we describe neutrino-electron scattering cross sections at tree and one loop level and define the running of the mixing angle used throughout the discussion. 
In Sec.~\ref{sec:thetaW_in_SNBD}, we outline the main assumptions regarding measuring the mixing angle at SBND and the statistical procedure applied. 
In Sec.~\ref{sec:results}, we present the results of our analysis, and finally, we draw our conclusions in Sec.~\ref{sec:conclusions}.

\section{Neutrino electron elastic scattering}
\label{sec:nu_e_scattering}

We measure the weak mixing angle by using neutrino-electron scattering events, $\nu+e\rightarrow \nu+e$.
This process probes $\sin^2\theta_W$ at a scale of the momentum transfer $Q^2 = 2 m_e T$, where $T$ and $m_e$ are the kinetic energy and mass of the recoil electron, respectively.
Because the BNB peaks around 1~GeV neutrino energy, this allows for the momentum transfer $Q \lesssim 20\ \rm MeV$, enabling a low-energy measurement of the weak mixing angle. 

The tree-level cross section for the scattering of a neutrino of flavor $\alpha$ off an electron at rest is given by 
\begin{equation}
    \dv{\sigma}{T} = \frac{2 G_F^2 m_e}{\pi} \left[g_L^2 + g_R^2\left(1 - \frac{T}{E_\nu} \right)^2 -g_L g_R \frac{m_e T}{E^2_\nu}\right],
    \label{eq:tree_level_xsec}
\end{equation}
where the flavor dependence of the cross section is encoded in the parameters $g_L$ and $g_R$, namely 
\begin{align}
    \begin{split}
        g_L& =\begin{dcases}
        -\frac{1}{2} - \sin^2{\theta_W} & \text{for } \nu_e,\\
        \frac{1}{2} - \sin^2{\theta_W} & \text{for } \nu_\mu,\nu_\tau,\\
        \end{dcases} \\
        g_R &= - \sin^2{\theta_W} \qquad\quad\; \ \text{for } \nu_e,\nu_\mu, \nu_\tau.\\
    \end{split}
    \label{eq:tree_level_couplings}
\end{align}
One-loop corrections to the tree level process introduce running of $\sin^2{\theta_W}$ and modify Eq.~\eqref{eq:tree_level_xsec} to~\cite{Bahcall:1995mm, Miranda:2021mqb}
\begin{align}
    \begin{split}
           \dv{\sigma}{T} &= \frac{2G_F^2 m_e}{\pi}\left[ \hat g_L^2\eta_{-}
           \! + \hat g_R^2\eta_{+}\!\left(1 - \frac{T}{E_\nu}\right)^2\!- \hat g_L\hat g_R\eta_{\pm}\frac{m_e T}{E^2_\nu}\right],
    \end{split}
    \label{eq:1_loop_xsec}
\end{align}
where $\hat g_{L,R}$ are now given by
\begin{align}
    \begin{split}
        \hat{g}_{L}& =\begin{dcases}
        \rho_{\rm NC} \left[-\frac{1}{2} - \hat{\kappa}^\ell(Q^2,\mu) \sin^2{\hat{\theta}_W}(\mu) \right] & \ell=e,\\
        \rho_{\rm NC} \left[ \frac{1}{2} - \hat{\kappa}^\ell(Q^2,\mu) \sin^2{\hat{\theta}_W}(\mu) \right] & \ell= \mu,\tau,\\
        \end{dcases} \\
        \hat{g}_{R} &= -\rho_{\rm NC} \hat{\kappa}^{\ell}(Q^2,\mu) \sin^2{\hat{\theta}_W}(\mu)
         \qquad\qquad\;\ \ \ell=e,\mu,\tau.
    \end{split}
    \label{eq:tree_level_couplings}
\end{align}
The corresponding antineutrino cross section can be obtained from Eq.~\eqref{eq:1_loop_xsec} by the replacement $\hat g_L \leftrightarrow \hat g_R$.
All quantities are computed within the $\overline{\rm MS}$ renormalization scheme. 
The loop corrections are encoded in three main parameters: the Fermi coupling constant $G_F$, used to absorb the majority of charged-current loop corrections~\cite{PhysRevD.22.2695} and fixed from muon decays, the parameter $\rho_{\rm NC} = 1.014032$, which is a flavor-independent correction to neutral-current processes~\cite{PhysRevD.22.2695, Ferroglia:2003wa, Marciano:2003eq}, and $\hat{\kappa}^{\ell}(Q^2,\mu)$ that encodes the flavor-dependent part of loop corrections to processes mediated by neutral currents~\cite{PhysRevD.22.2695,Sarantakos:1982bp, Ferroglia:2003wa, Marciano:2003eq}. The functions $\eta_+$, $\eta_-$, and $\eta_{\pm}$ account for the QED corrections and can be found in Appendix~\ref{ap:QED_corrections}. 
The cross section, when expressed in terms of these combinations, becomes finite as all divergences cancel.

While $\sin^2\hat\theta_W(\mu)$  can be used for electroweak precision measurements, it is a function of the renormalization scale $\mu$ (the t'Hooft parameter in dimensional regularization), which is not a physical parameter.
In the literature, it is common to define a weak mixing angle as a function of the momentum transfer, which is measured experimentally.
To understand how this is done, we note that although both $\hat\kappa^\ell$ and $\sin^2\hat\theta_W$ depend on the scale $\mu$, their product is $\mu$-independent~\cite{Marciano:2003eq, Sirlin:2012mh, Sarantakos:1982bp, Bahcall:1995mm, Miranda:2021mqb, Kumar:2013yoa, Ferroglia:2003wa}. 
This means we can choose any value of $\mu$, and we do so in a way that simplifies the expressions for these quantities. 
Specifically, we choose $\mu = m_Z$. 
From here on, we will adopt the common practice in the literature and define $\hat\kappa^\ell(Q^2) \equiv \hat\kappa^\ell(Q^2, m_Z)$.
Note that, at this point, we could define a flavor-dependent weak mixing angle: the product $\hat\kappa^\ell(Q^2)\sin^2\hat\theta_W(m_Z)$, but it would be inconvenient to compare different experimental results.

A flavor-independent definition of the weak mixing angle is accomplished by extracting the flavor-independent part of  $\hat{\kappa}^\ell(Q^2)$ into a new parameter $\hat\kappa(Q^2)$~\cite{Ferroglia:2003wa, Erler:2004in}. 
We follow Ref.~\cite{Erler:2004in}, which introduced the effective weak mixing angle $\sin^2\theta_W^{\rm eff}(Q^2)$, such that at $Q^2=0$, we have
\begin{equation}
    \sin^2{\hat{\theta}_W^{\rm eff}}(Q^2=0) = \hat{\kappa}(0) \sin^2{\hat{\theta}_W}(m_Z),
    \label{eq:run_theta_weak}
\end{equation}
where 
\begin{equation}
    \hat{\kappa}(0) = 1.03232 \pm 0.00029,
\end{equation}
is a process-independent correction. 

In a nutshell, to extract the flavor-independent weak mixing angle, one needs to measure the neutrino-electron cross section experimentally, fit the value of $\sin^2\hat\theta_W(\mu=m_Z)$ using \cref{eq:1_loop_xsec}, and infer the effective weak mixing at low scales with \cref{eq:run_theta_weak}.
By using $\sin^2\hat\theta_W(\mu = m_Z)=0.23129$ derived from collider data, including LEP, we obtain $\sin^2{\hat{\theta}_W^{\rm eff}}(0) = 0.239$ as the SM prediction for the effective weak mixing angle at SBND. 
To avoid cluttering, we will always refer to the effective weak mixing angle as the weak mixing angle, and we will use this quantity in all our plots and analyses.

\section{The weak mixing angle at SBND}
\label{sec:thetaW_in_SNBD}

\begin{figure}[t]
    \includegraphics[width=\columnwidth]{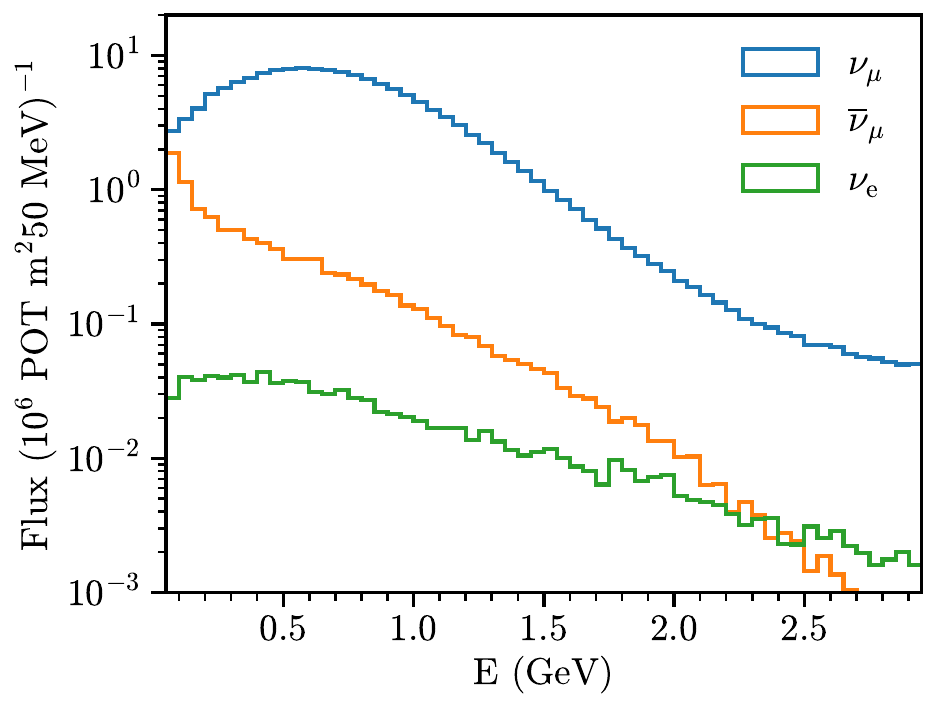}
    \caption{Volume-integrated flux at SBND, from Ref.~\cite{slides}. We do not show the subdominant $\bar \nu_e$ component.}
    \label{fig:flux_plot}
\end{figure} 

The SBND experiment is a liquid argon time projection chamber (LArTPC) neutrino detector downstream Fermilab's BNB line~\cite{MicroBooNE:2015bmn, Machado:2019oxb}.
SBND is located 110~m from the beam target. Its proximity to the target will allow it to collect unprecedented statistics. 
Figure~\ref{fig:flux_plot} shows the neutrino fluxes in the detector~\cite{slides}.
The flux is dominated by the $\nu_\mu$, which peaks at about 0.8~GeV.
There is a 5\% $\overline\nu_\mu$ contamination and a $0.5\%$ $\nu_e$ component. 
We note that in its proposal, there is no plan to run SBND in antineutrino mode~\cite{MicroBooNE:2015bmn}.

The rate of $\nu-e$ events depends on the flux composition through the flavor dependence of the cross section. 
Neutrino- and antineutrino-electron scattering cross sections exhibit a different dependence on the weak mixing angle, specifically $\hat g_L\leftrightarrow \hat g_R$ as $\nu_\ell\leftrightarrow\bar\nu_\ell$ in Eq.~\eqref{eq:1_loop_xsec}. 
The sensitivity of SBND to $\sin^2{\theta_W}$ is, therefore, boosted by the high statistics it will collect and affected by the flavor dependence of the cross section. 

\subsection{Event simulation and experimental cuts}

For the mock analysis, we generate signal events in the following way.
We implement a Monte Carlo to generate the neutrino-electron scattering signal using the next-to-leading order cross section in~\cref{eq:1_loop_xsec}. 
Our signal corresponds to a single electromagnetic (EM) shower produced by the recoiled electron. We then apply detector smearing of the outgoing electrons with an angular resolution of $2^\circ$. This provides a $E_{e}\theta_e^2$ spectrum in agreement with Ref.~\cite{slidesnuescattering}, where $E_{e}$ and $\theta_e$ are the outgoing electron's energy and angle with respect to the beam direction. We neglect energy smearing of the electron in our analysis because its effect is subdominant compared to that of angular smearing in the $E_{e}\theta_e^2$ distribution.

The dominant background events come from neutrino-nucleus scattering. These events can mimic our signal if the only visible final state particle---the particle above the energy threshold---is either an electron or a photon. Both of these particles would generate EM showers in the TPC. 
We simulate the background using \texttt{NuWro}~\cite{GOLAN2012499, Juszczak:2005zs}.

Although the neutrino-nucleus event rate is over 1,000 times larger than that of the neutrino-electron signal, we benefit from several features of the latter to improve the signal-to-background ratio. 
Outgoing electrons in neutrino-electron scattering are very forward.
Kinematics constrain $E_{e}\theta_e^2 < 2m_e$. 
This is a powerful cut that rejects most of the neutrino-nucleus background.
Besides, neutrino-electron scattering events do not involve hadrons, so cutting on any visible hadronic activity can also strongly suppress backgrounds.

Precisely, we mock detector responses by assuming energy thresholds of 30~MeV for protons, 500~MeV for neutrons, and 15~MeV for electrons and photons.
We decay $\pi^0\to\gamma\gamma$ before applying cuts.
Events with any visible hadronic activity are rejected.
Events with more than one EM shower can also be rejected, but we need to estimate the number of events with distinguishable EM showers.
To that end, we also smear the direction of outgoing photons by $2^\circ$ and cut events with two EM showers with opening angles above $3^\circ$.
In principle, the experiment could also cut on visible displaced EM vertices, but this would be challenging for showers with small angular separation. Therefore we do not consider this possibility here.
We then apply a cut  $E_e\theta_e^2<1.5$~MeV, which further reduces backgrounds.

\begin{figure}[t]
    \includegraphics[width=0.98\columnwidth]{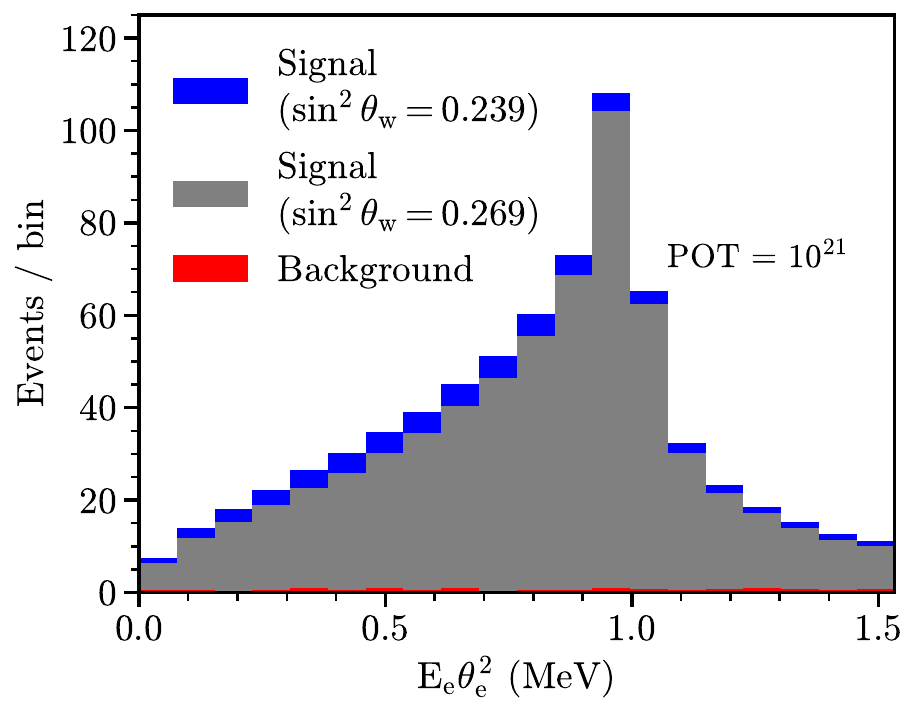}
    \vspace{-0.1em}
    \caption{SBND's $E_e\theta_e^2$ spectra for neutrino-electron signal for two different $\sin^2{\theta_W}$ values (blue and grey) and neutrino-nucleus background events (red) after cuts.}
    \label{fig:signal_bkg_ratio}
\end{figure}

In Fig.~\ref{fig:signal_bkg_ratio}, we present the $E_e\theta_e^2$ spectrum for signal events: blue for the SM value $\sin^2{\theta_W} = 0.239$, gray for $\sin^2{\theta_W} = 0.269$, and red for the background. These are expected for $10^{21}$~protons on target (POT). We note that varying the weak mixing angle changes both the shape and the normalization of the events spectrum, and the cuts lead to essentially a background-free neutrino-electron sample while still retaining most signal events.

\subsection{SBND-PRISM}

\begin{figure}[t]
    \includegraphics[width=\columnwidth]{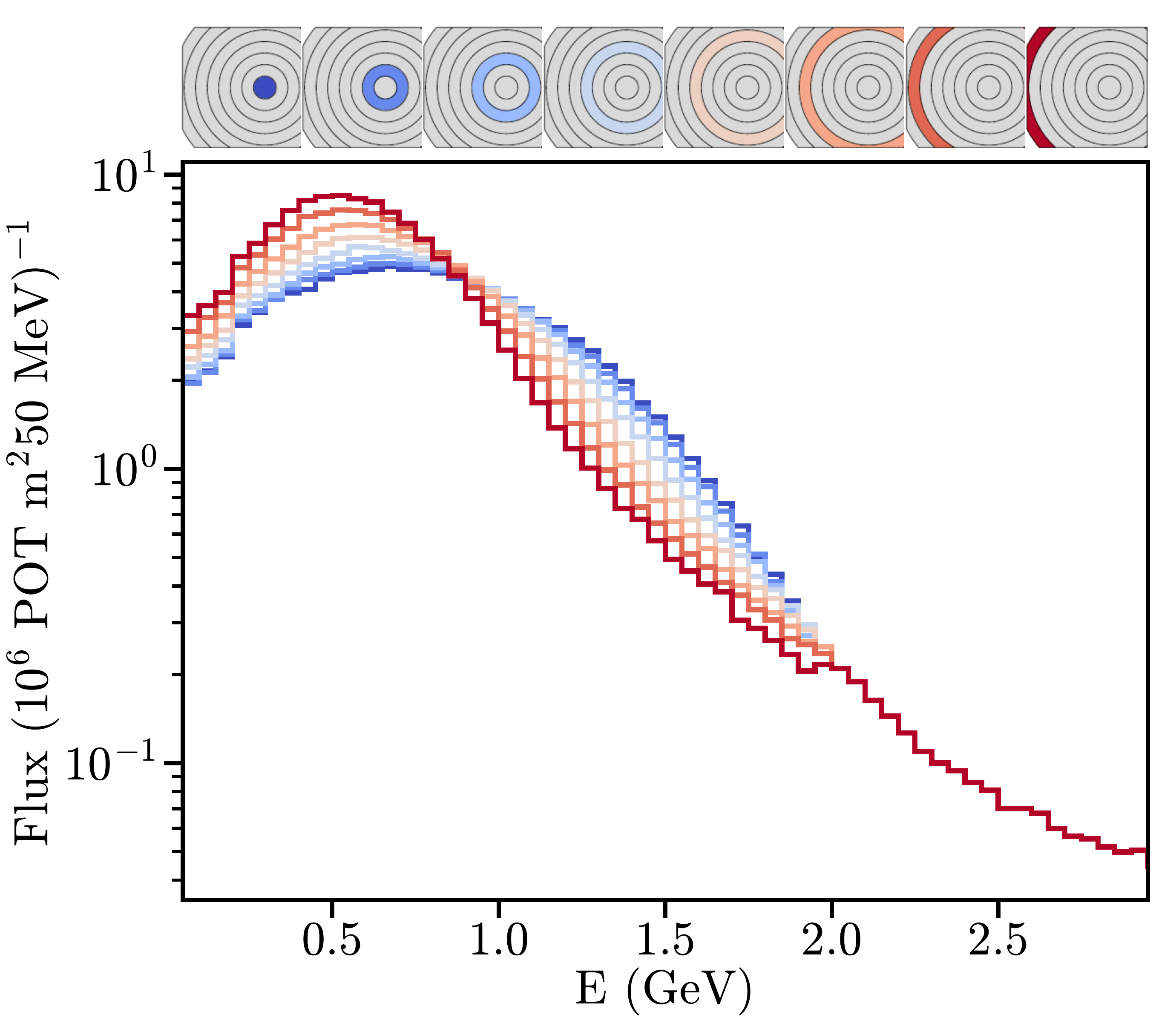}
    \caption{Layer-dependent $\nu_\mu$ flux from Ref.~\cite{slides}. The top panel depicts the front view of the SBND experiment and the eight slices associated with the different layers (adapted from Ref.~\cite{Alves:2024djc}). The main panel shows the fluxes in each layer, with the color matching that of the corresponding layer.}
    \label{fig:numu_flux_SBND_PRISM}
\end{figure}

The PRISM concept is a potential way to mitigate systematic uncertainties related to the neutrino flux, possibly improving the sensitivity to $\sin^2{\theta_W}$.
It utilizes the fact that neutrino fluxes have a geometrical dependence, meaning that there is a nontrivial correlation between the outgoing neutrino angle relative to the beam axis and the neutrino energy profile~\cite{delTutto2021SBND} (see also Refs.~\cite{nuprism, dunetdr}). Consequently, different detector regions see different neutrino fluxes with varying spectra.
This technique provides more opportunities to use the recoil electron's spectral information. The proximity of the detector to the neutrino source makes SBND especially suitable for utilizing the PRISM technique.

\begin{figure}[t]
    \includegraphics[width=\columnwidth]{"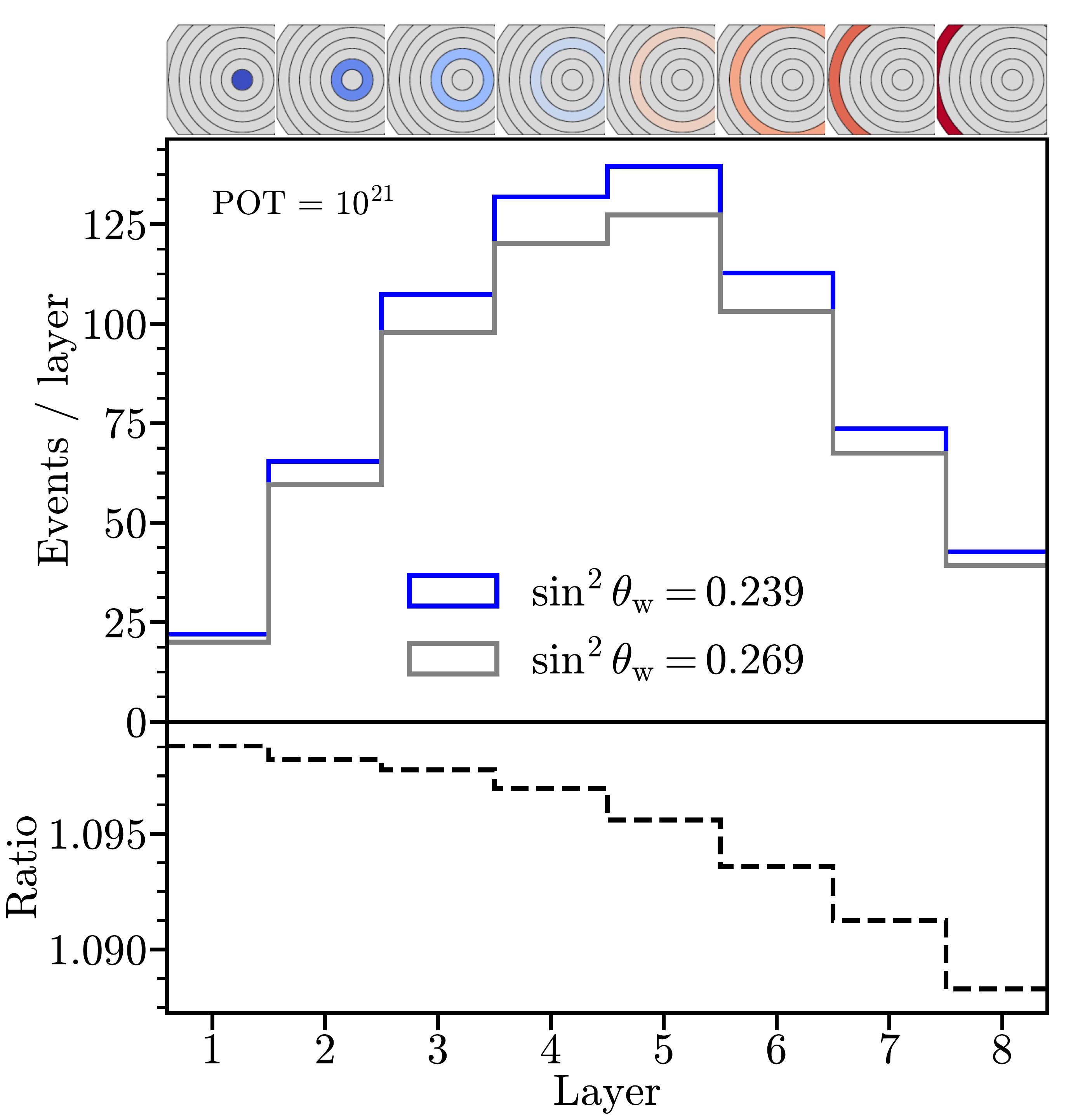"}
    \caption{{\bf Top:} Number of neutrino-electron scattering events in each SBND-PRISM layer for two values of $\sin^2\theta_W$. 
    {\bf Bottom:} Ratio between the number of events for the two values of $\sin^2\theta_W$ in each layer.}
    \label{fig:event_ratio}
\end{figure}

In greater detail, the SBND detector is logically divided into 8 concentric layers around the beam axis. Figure~\ref{fig:numu_flux_SBND_PRISM} displays the $\nu_\mu$ flux across different layers for SBND-PRISM from Ref.~\cite{slides}, with the layers depicted by the colored circles above the plot (adapted from Ref.~\cite{Alves:2024djc}). 
The most important feature of the flux that changes across different layers is the peak energy. 
It decreases as we move to the outer layers, effectively lowering the flux-averaged cross section and hence impacting the total count of events in each layer. 
Figure~\ref{fig:event_ratio} shows the number of neutrino-electron scattering events in each SBND-PRISM layer for the planned exposure of $10^{21}$~POT, for two benchmark values of $\sin^2\theta_W$. 
The number of events is computed similarly to SBND, but using the neutrino flux for each layer accordingly.
The larger number of events in the middle layers is a volume effect.
The lower panel shows the ratio of events between these benchmarks.
The ratio changing as a function of the layer indicates that utilizing signal differences in different layers should, in principle, better constrain $\sin^2{\theta_W}$. However, the mild layer dependence also suggests that the advantage of PRISM will be modest. 
In addition, the PRISM technique could also help improve $\sin^2{\theta_W}$ sensitivity because it can mitigate the overall flux uncertainty.

To implement PRISM in our analysis, we need the neutrino fluxes of all flavors in each layer. This information is not public yet. 
Reference~\cite{slides} provides the ratio of $\nu_e+\overline\nu_e$ to $\nu_\mu+\overline\nu_\mu$ scattering events on argon, for each off-axis position. We detail in Appendix~\ref{ap:PRISM_fluxes} two methods we developed to obtain the missing fluxes from this ratio and the public SBND fluxes. Nevertheless, as we will discuss later, our analysis shows that for realistic exposures, there will be little advantage in using the PRISM concept, and therefore, our results do not depend on the way we derive the missing fluxes.
     
\subsection{Statistical analysis}

\begin{figure}[t]
        \includegraphics[width=\columnwidth]{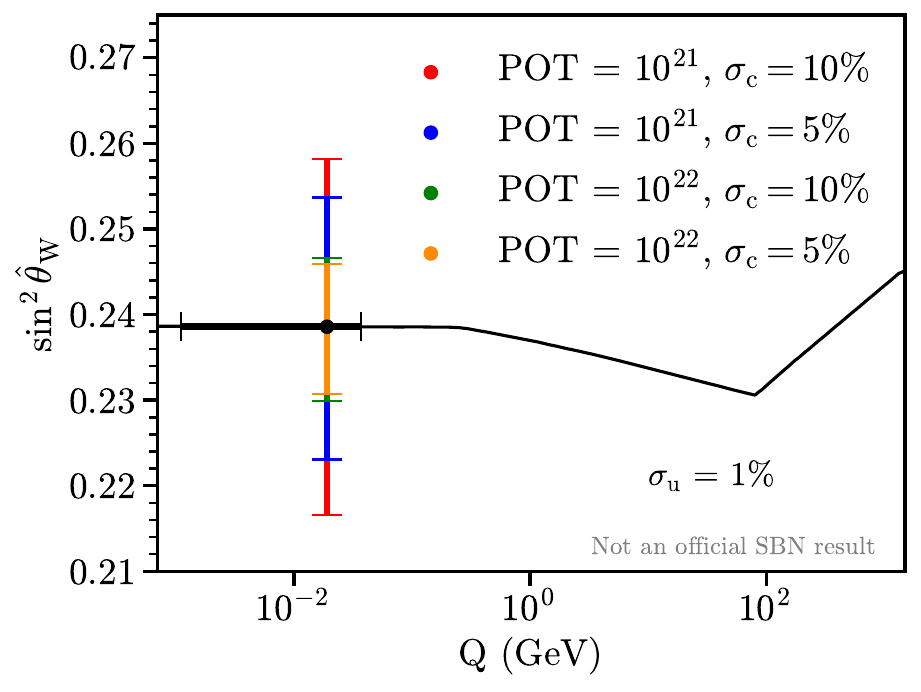}
        \caption{The sensitivity of SBND considering multiple benchmark values for exposure and correlated systematic uncertainties. The black line represents the mixing angle running as expected from the Standard Model. 
        We denote the correlated and uncorrelated flux uncertainties by $\sigma_c$ and $\sigma_u$, respectively.}
        \label{fig:final_fit}
\end{figure}

For the statistical analysis, we define a Gaussian log-likelihood ratio  as our test statistic,
\begin{equation}
    \chi^2 = \Delta^T \Sigma^{-1} \Delta\,,
\end{equation}
where $\Delta = N^{\rm obs} - N^{\rm pred}$ is a vector of the difference between mock data $N^{\rm obs}$ and theory prediction $N^{\rm pred}$, containing 20 bins in the $E_e\theta_e^2$ spectrum for each of the 8 layers (i.e. a $160 = 8 \times 20$ dimensional vector). $\Sigma$ is a simplified covariance matrix that we assume to be of the form
\begin{equation}
    \Sigma_{ij} = (\sigma_{c}^2 + \delta_{ij}\sigma_{u}^2)N^{\rm pred}_i N^{\rm pred}_j + \delta_{ij}N^{\rm pred}_j\,,
\end{equation}
where $\sigma_{u}$ and $\sigma_{c}$ refer to uncorrelated and correlated systematic uncertainties, respectively. 

One of the main goals of our analysis is to evaluate how statistics, systematic errors, and the PRISM technique interplay with each other and what contributes the most to the $\sin^2\theta_W$ sensitivity. We will study two benchmark scenarios regarding the exposure: the planned exposure of $10^{21}$~POT and the optimistic scenario with $10^{22}$~POT. We choose the uncorrelated systematic uncertainties to be $\sigma_{u} = 1\%$. As for $\sigma_{c}$, we also study two benchmark cases of 5\% and 10\%, where the latter value is motivated by the official MicroBooNE flux uncertainties~\cite{MicroBooNE:2018efi}.

\section{Results}
\label{sec:results}

\begin{figure}[t]
    \includegraphics[width=\columnwidth]{"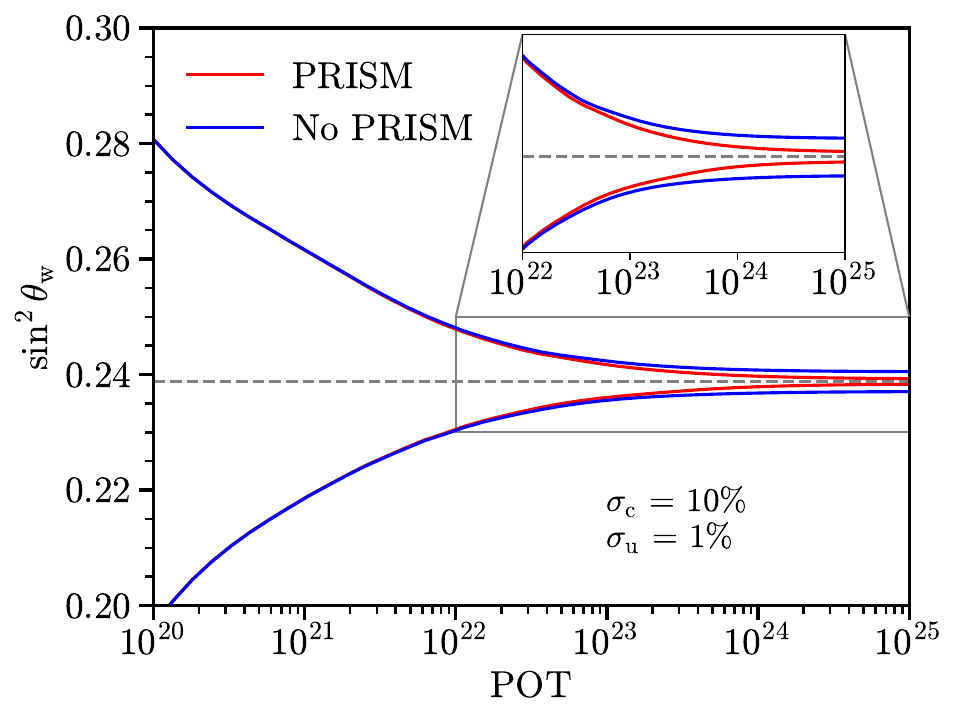"}
    \caption{SBND's uncertainty on the weak mixing angle as a function of the exposure with and without SBND-PRISM.}
    \label{fig:PRISM_effect_POT}
\end{figure}

In Figure~\ref{fig:final_fit}, we present the estimated SBND sensitivity to $\sin^2\theta_W$ for various benchmark scenarios. 
The SM prediction for the scale dependence of $\sin^2\theta_W$ is represented by a black line.
For the nominal $10^{21}$ POT exposure, a conservative 10\% correlated systematics would yield a $\sin^2 \theta_W$ sensitivity of 8\%.
The measurement is primarily driven by the total event count and the shape of the $E_e \theta_e^2$ spectrum. 
The sensitivity is mainly limited by the correlated uncertainty $\sigma_c$.
An improvement of systematics to 5\% would improve the sensitivity to 6\%.
Meanwhile, if SBND could increase the exposure, e.g., to $10^{22}$ POT, there would be a much more significant improvement to $\sin^2 \theta_W$ sensitivity, to 3\%, which comes from better utilization of the spectral shape information.

To see how the PRISM technique impacts the determination of $\sin^2\theta_W$, we show in Fig.~\ref{fig:PRISM_effect_POT} the evolution of the sensitivities with exposure, with and without PRISM. For the realistic exposure of $10^{21}$~POT, PRISM offers little advantage.  Each PRISM layer contains only about $\mathcal{O}(100)$ events or fewer, providing insufficient additional information from the layer-dependent $E_e \theta_e^2$ spectrum. However, for an exposure of $10^{22}$~POT or higher, the PRISM concept can enhance the sensitivity by up to a factor of two, effectively leveraging systematic uncertainties through the additional layer-dependent $E\theta_e^2$ information. While reaching this exposure would require an extended data-taking period, the upcoming upgrade to the Fermilab accelerator complex, particularly the BNB, may allow for a higher annual POT delivery~\cite{Gori:2024zbs}. 

\begin{figure}[t]
        \includegraphics[width=\columnwidth]{"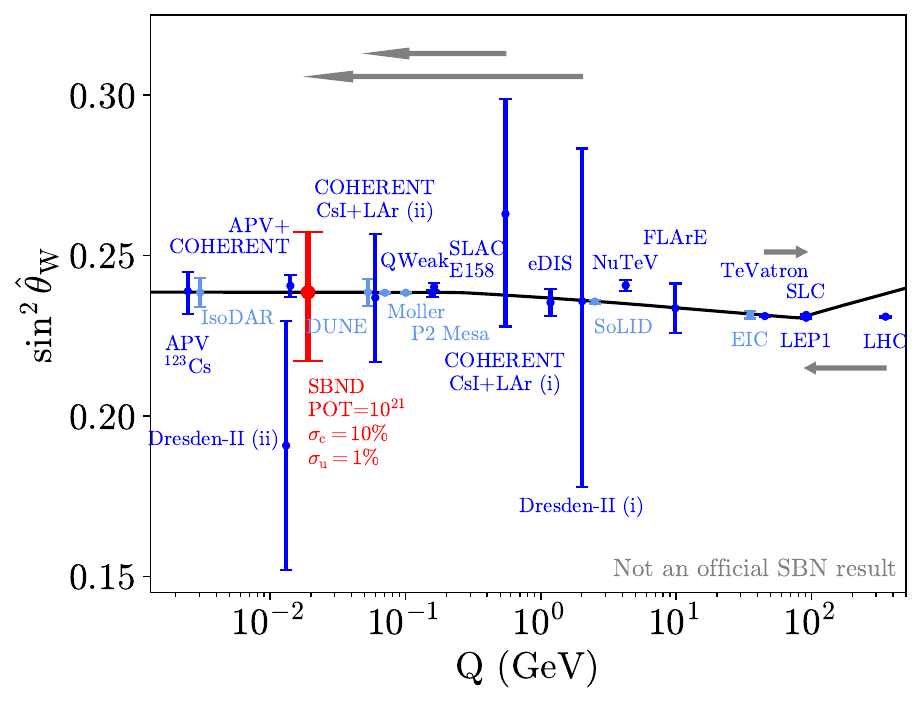"}
        \caption{Sensitivity to the weak mixing angle for SBND compared to other measurements~\cite{ParticleDataGroup:2022pth, Cadeddu:2021ijh, deGouvea:2019wav, Cadeddu:2020lky, DeRomeri:2022twg, Majumdar:2022nby, AristizabalSierra:2022axl, MammenAbraham:2023psg, Boughezal:2022pmb,2014arXiv1411.4088M,Berger:2015aaa, Zhao:2017xej, Alonso:2021kyu}. 
        Dark blue points illustrate existing measurements and light blue projection. 
        We displace COHERENT CsI+LAr (i),  Dresden-II (i) for visibility, TeVatron, and LHC for visibility.        
        \label{fig:weak_mixing_angle_final}}
\end{figure}

Figure~\ref{fig:weak_mixing_angle_final} shows how SBND sensitivity compares with other experiments~\cite{ParticleDataGroup:2022pth, Cadeddu:2021ijh,deGouvea:2019wav, Cadeddu:2020lky, DeRomeri:2022twg, Majumdar:2022nby, AristizabalSierra:2022axl, MammenAbraham:2023psg, Boughezal:2022pmb,2014arXiv1411.4088M,Berger:2015aaa, Zhao:2017xej, Alonso:2021kyu}. We can see that SBND has the potential to be the most competitive neutrino measurement of the weak mixing angle below the GeV scale until DUNE takes sufficient data. Only then will neutrino measurements be competitive with other low-energy methods, such as atomic parity violation.

\section{Conclusions}
\label{sec:conclusions}

We have estimated SBND's sensitivity to the weak mixing angle $\sin^2\theta_W$, at the $Q=20$~MeV scale, by measuring neutrino-electron scattering. 
We performed a detailed analysis, implementing reconstruction and cuts on an event-by-event basis.
Cutting events with $E_e\theta_e^2$ above 1.5~MeV, any visible amount of hadronic activity, and more than one visible electromagnetic shower allows for a virtually background-free search.

Our results show that SBND can place a strong constraint on $\sin^2\theta_W$, competitive with current and future neutrino experiments.
Nevertheless, observing running from $m_Z$ down to 20~MeV would require a longer-than-planned exposure of order $10^{22}$ POT.
Even though there is no concrete plan to reach such an exposure, Fermilab's Accelerator Complex Evolution plan could pave the way to this possibility.
We have also studied the relevance of SBND-PRISM to this measurement and shown that it is more relevant for larger exposures.

\begin{acknowledgments}

Fermilab is managed by the Fermi Research Alliance, LLC (FRA), acting under Contract No. DE-AC02-07CH11359.
Y.F.P.G.~is supported by the UK Science and Technology Facilities Council (STFC) under grant ST/T001011/1. Both G.F.S.A. and A.P.F. received full financial support from the São Paulo Research Foundation (FAPESP) through the following grants: G.F.S.A. under Contracts No. 2022/10894-8 and No. 2020/08096-0, and A.P.F. under Contracts No. 2021/03496-3 and No. 2022/10057-9. G.F.S.A. and A. P. F. would like to thank the Fermilab Theory Group for its hospitality.

\end{acknowledgments}

\bibliographystyle{JHEP}
\bibliography{refs}

\appendix

\section{QED corrections to the neutrino electron elastic scattering}
\label{ap:QED_corrections}

In this Appendix, we provide the expressions and values of quantities defined in \cref{eq:1_loop_xsec,eq:tree_level_couplings}.
The process-dependent coefficients relevant to our analysis are given by~\cite{Bahcall:1995mm}
\begin{align}
        \hat{\kappa}^{e}(T) &= 0.9791 + 0.0097 I(T) \pm 0.0025,\label{eq:kappa_nue}\\
        \hat{\kappa}^{\mu}(T)&= 0.9970 - 0.00037 I(T) \pm 0.0025,
    \label{eq:kappa_numu}
\end{align}
where $T$ is the electron recoil,
\begin{equation}
    I(T) = \frac{1}{6}\left\{\frac{1}{3} + (3-x^2)\left[\frac{x}{2}\log\left(\frac{x+1}{x-1}\right) \right] \right\},
\end{equation}
and $x = \sqrt{1 + 4m_e/T}$.

The QED correction functions~\cite{Bahcall:1995mm, Miranda:2021mqb} are given by $\eta_{+} = 1 + (\alpha/\pi)f_{+}$, $\eta_{-} = 1 + (\alpha/\pi)f_{-}$ and $\eta_{\pm} = 1 + (\alpha/\pi)f_{\pm}$.
The functions $f_+$, $f_-$ and $f_\pm$ are defined as
\begin{widetext}
   \begin{align}
	   f_{+}(z) &= \frac{1}{ \left(1-z\right)^2}\left[ \frac{E_e}{l}\ln\left( \frac{E_e+l}{m_e}\right)-1\right]\left\lbrace \left( 1-z\right)^2 \left[2\ln\left( 1 - z -\frac{m_e}{E_e+l}\right)\right. \right. \nonumber \left. \left. -\ln\left( 1-z \right) -\frac{1}{2}\ln z - \frac{2}{3}\right] - \frac{z^2\ln z + 1-z}{2} \right\rbrace \nonumber\\
            &-  \frac{\left( 1-z\right)^2}{2}\left\lbrace \ln^2 \left( 1-z\right) + \beta \left[ L(1-z) - \ln z \ln \left( 1-z\right) \right] \right\rbrace  \nonumber + \ln \left( 1-z\right) \left[ \frac{z^2}{2}\ln z + \frac{1-z}{3} \left( 2z-\frac{1}{2} \right)  \right] -\frac{z^2}{2} L\left(1-z \right)  \nonumber\\
            &\qquad {} - \frac{z \left( 1-2z\right) }{3} \ln z - \frac{z\left( 1-z\right)}{6} - \frac{\beta}{12}\left[\ln z + \left( 1-z\right)\left( \frac{115-109z}{6}\right) \right] \,,\\
            f_{-}(z)&= \left[\frac{E_e}{l}\ln\left( \frac{E_e+l}{m_e}\right)-1 \right] \left[ 2\ln\left( 1 - z -\frac{m_e}{E_e+l} \right) -\ln\left( 1-z \right) -\frac{1}{2}\ln z - \frac{5}{12}\right] +  \frac{1}{2}\left[ L(z) - L(\beta) \right]\nonumber\\
            &- \frac{1}{2}\ln^2\left( 1-z \right) -\left( \frac{11}{12} + \frac{z}{2}\right) \ln \left( 1-z \right) +  z\left[\ln z + \frac{1}{2}\ln\left( \frac{2E_{\nu}}{m_e}\right)\right] - \left( \frac{31}{18} + \frac{1}{12}\ln z \right)\beta -\frac{11}{12}z + \frac{z^2}{24} \,,\\
            f_{\pm}(z) &= \left[\frac{E_e}{l}\ln{\biggl(\frac{E_e + l}{m_e} \biggr)} \right]2\ln{\biggl(1 - z - \frac{m_e}{E_e + l} \biggr)},
            \label{eq:QED_functions}
    \end{align}
\end{widetext}
where $z = T/E_{\nu}$, $E_e = T + m_e$, $l = \sqrt{E_e^2 - m_e^2}$, $\beta = l/E_e$ and $L(x)$ is defined as
\begin{equation}
    L(x) =  \int_0^x \frac{\ln{|1-t|}}{t}dt.
\end{equation}

\section{Details of SBND-PRISM fluxes}
\label{ap:PRISM_fluxes}

\subsection{SBND-PRISM fluxes}

\begin{figure}[t]
    \includegraphics[width=\columnwidth]{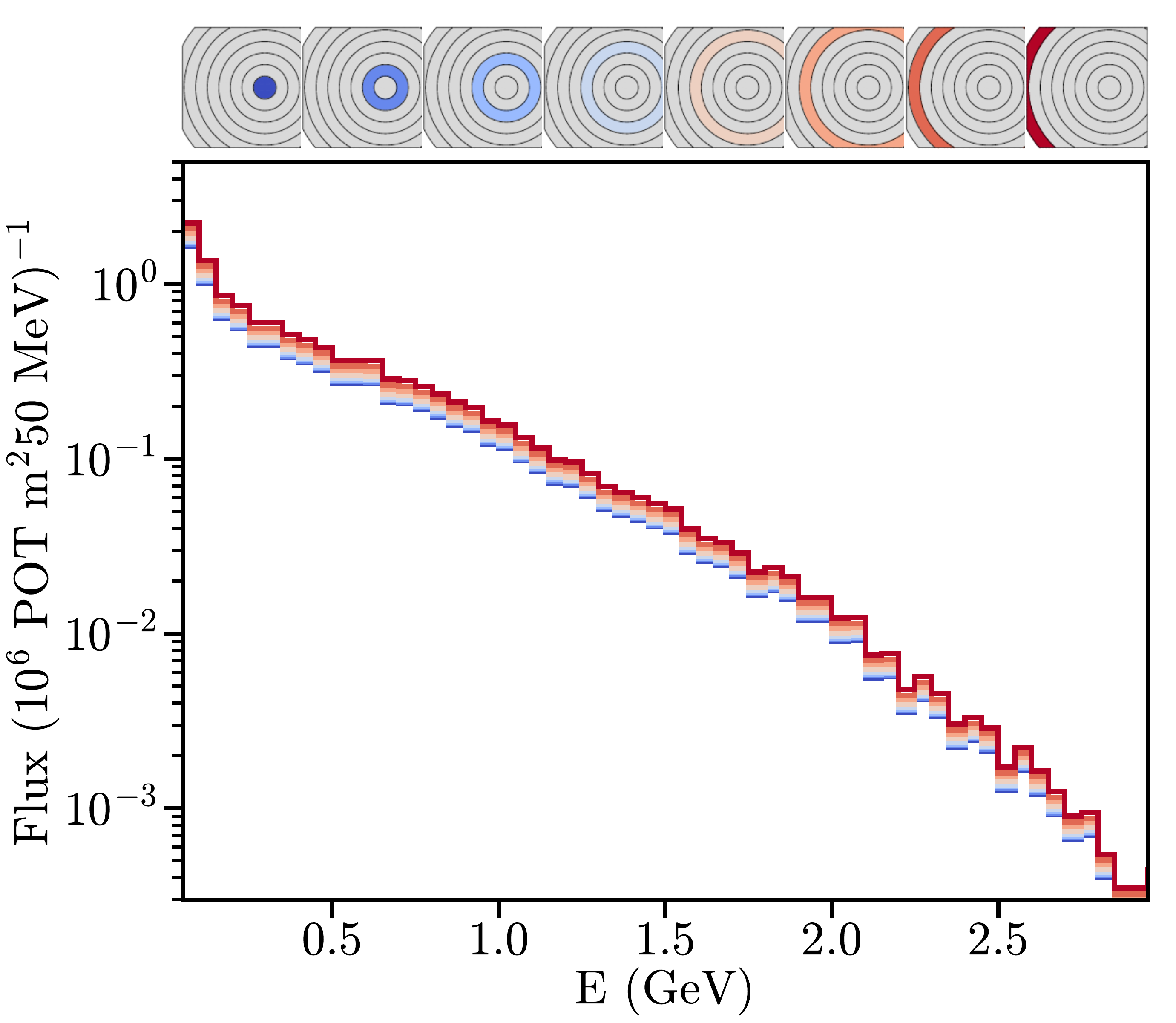}
    \caption{Our estimate for the layer-dependent $\bar \nu_\mu$ flux, considering the approach from~\cref{eq:case-1-numubar}. The color coding corresponds to the layer associated with the flux.}
    \label{fig:numubar_flux_SBND_PRISM}
\end{figure}

\begin{figure}[t]
        \includegraphics[width=\columnwidth]{"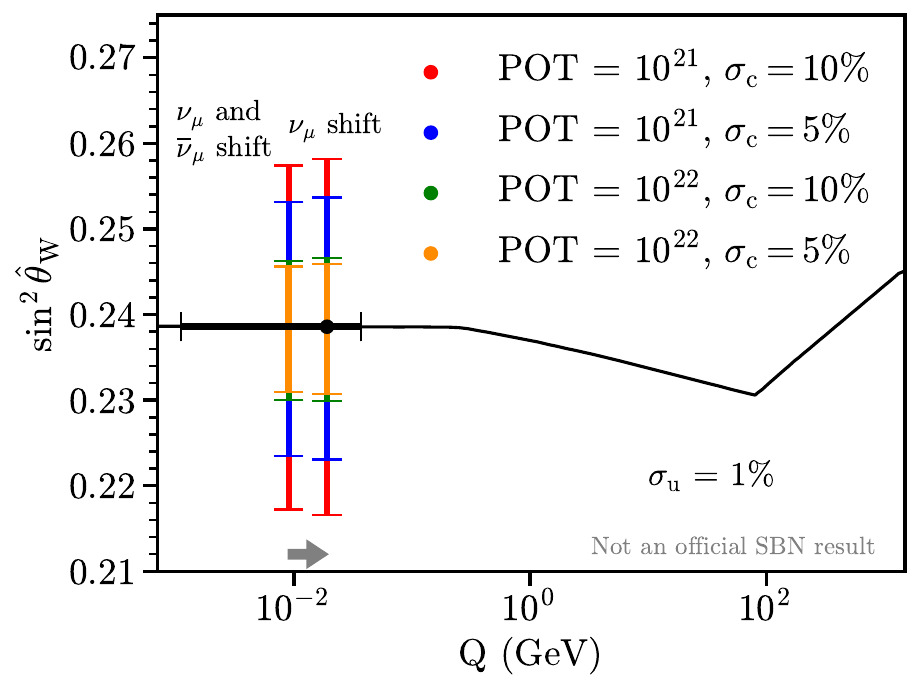"}
        \caption{Comparison of SBND sensitivities using two different approaches to extract neutrino flux/composition in each layer. The $\nu_\mu$ shift ($\nu_\mu$ and $\overline{\nu}_\mu$ shift) assumption corresponds to case 1 (case 2) in the Appendix. The result for case 2 is artificially dislocated to the left-hand side.}
        \label{fig:fit_with_all_flavors}
\end{figure}

To estimate SBND's sensitivity to $\sin^2\theta_W$, we need the flavor composition of the neutrino flux at each off-axis layer, specifically $\nu_\mu$, $\overline\nu_\mu$, and $\nu_e$. 
Reference~\cite{slides} presents the layer-dependent $\nu_\mu$ flux and the ratio of $\nu_e+\overline\nu_e$ to $\nu_\mu+\overline\nu_\mu$ scattering events on argon in each layer. 
Neglecting the subdominant $\overline{\nu}_e$ contamination and assuming that the $\nu_\mu$-Ar and $\nu_e$-Ar cross sections are approximately the same, we can express this ratio $R_i$ as
\begin{equation}
\label{eq:ratio_flavors}
    R_i \simeq \frac{\phi_e^i}{\phi_\mu^i+\phi_{\bar\mu}^i},
\end{equation}
where $i$ denotes the SBND-PRISM layer. Consequently, we need to infer the $\nu_e$ and $\overline\nu_\mu$ fluxes in each layer from $\nu_\mu$ fluxes and $R_i$.

We first make the assumption that the $\nu_e$ flux shape is the same in each layer and determined by the SBND $\nu_e$ flux, i.e., the total $\nu_e$ flux without accounting for any SBND-PRISM effect. In other words, $\phi_e^i=N_i\phi_e^{\rm SBND}$. The normalization factor $N_i$ can be easily found by imposing the correct number of $\nu_e$ events in each layer as presented in Ref.~\cite{slides}.
This layer-independent assumption is reasonable: because $\nu_e$ comes mainly from three-body kaon decays, their spectra exhibit only a mild dependence with the off-axis angle~\cite{slidesnuescattering}. 

We test two methods to infer the $\overline \nu_\mu$ component. In the first case, we assume it behaves like the $\nu_e$, meaning it remains constant across different layers and is determined by the total $\overline \nu_\mu$ flux. The only adjustment comes from the layer volume effect, $\phi_{\bar\mu}^i=N_i\phi_{\bar\mu}^{\rm SBND}$. Figure~\ref{fig:numubar_flux_SBND_PRISM} shows the resulting $\overline \nu_\mu$ flux for this case. Here, any layer dependence in $R_i$ comes from the $\nu_\mu$ flux $\phi_\mu^i$. Since Ref.~\cite{slidesnuescattering} provides the normalized $\nu_\mu$ flux in each layer, which we denote as $\tilde\phi_\mu^i$, we take $\phi_\mu^i = a_i N_i\tilde\phi_\mu^i$. 
The layer-dependent constants $a_i$ can be determined from $R_i$ as
\begin{equation}
\label{eq:case-1-numubar}
    \text{Case 1:}\qquad R_i = \frac{\phi_e^{\rm SBND}}{a_i\tilde\phi_\mu^i+\phi_{\bar\mu}^{\rm SBND}}.
\end{equation}

An alternative way to extrapolate the $\bar\nu_\mu$ flux is the following. We assume that both $\nu_\mu$ and $\overline\nu_\mu$ fluxes change with off-axis angles by the same amount, that is, $\phi_\mu^i = b_i N_i\tilde\phi_\mu^i$ and $\phi_{\bar\mu}^i = b_i N_i\phi_{\bar\mu}^{\rm ff}$.
In this case, the constants $b_i$ are determined from
\begin{equation}
\label{eq:case-2-numubar}
    \text{Case 2:}\qquad R_i = \frac{\phi_e^{\rm SBND}}{b_i (\tilde\phi_\mu^i+\phi_{\bar\mu}^{\rm SBND})}.
\end{equation}
In reality, the muon antineutrino flux should behave somewhere in between the $\nu_\mu$ and $\nu_e$ fluxes. This is what happens, for example, to the  DUNE beam, although it is higher in energy than SBND. 

Figure~\ref{fig:fit_with_all_flavors} show the $\sin^2\theta_W$ sensitivities for cases 1 and 2 in~\cref{eq:case-1-numubar,eq:case-2-numubar}. 
Both are slightly displaced for better visibility. 
We see that there are only slight differences between the two cases.

\end{document}